\documentclass{PoS}

\usepackage{graphicx}
\usepackage{amsmath}
\usepackage{amssymb}

\newcommand{\simgt}%
{\mathrel{\lower0.5ex\hbox{\ensuremath{\buildrel>\over\sim}}}}
\newcommand{\simlt}%
{\mathrel{\lower0.5ex\hbox{\ensuremath{\buildrel<\over\sim}}}}
\newcommand{\BarLongRightArrow}%
{\ensuremath{\mathrel\vert\joinrel\Longrightarrow}}

\newcommand{\etal}{\textit{et al.}}

\newcommand{\be}{\begin{equation}}
\newcommand{\ee}{\end{equation}}

\title{Perturbative matching of heavy-light currents at one-loop}

\ShortTitle{Perturbative matching of heavy-light currents at one-loop}

\author{\speaker{Aida~X.~El-Khadra} and Elvira~G\'amiz \\
Department of Physics, University of Illinois, Urbana, IL 61801, USA \\
E-mail: \email{axk@uiuc.edu}, \email{megamiz@uiuc.edu}}

\author{Andreas~S.~Kronfeld \\
Theoretical Physics Department, Fermi National Accelerator Laboratory,
Batavia, IL 60510, USA \\
E-mail: \email{ask@fnal.gov}}

\author{Matthew~A.~Nobes \\
Department of Physics, Cornell University, Ithaca, NY 14853, USA \\
and \\
Brevan Howard Asset Management, London, SW1Y 6XA, United Kingdom \\
E-mail: \email{drnobes@gmail.com}}

\abstract{
We present results of a perturbative matching calculation
performed at one-loop for heavy-light currents. We use the
Fermilab action for the heavy quarks, the Asqtad action
for the light quarks, and an improved gluon action. We
also present results for heavy-heavy currents with Fermilab
heavy quarks and improved glue.
}

\FullConference{The XXV International Symposium on Lattice Field Theory\\
		 July 30-4 August 2007\\
		 Regensburg, Germany}

\begin{document}

\section{Introduction}

The Fermilab Lattice and MILC collaboration's program includes
calculations of the hadronic matrix elements for weak $D$ and
$B$ meson decays, in particular, the decay constants $f_D$, $f_{D_s}$,
$f_B$, and $f_{B_s}$ and the semileptonic form factors for
$B \rightarrow \pi \ell \nu$, $D \rightarrow \pi (K) \ell \nu$,
and $B \rightarrow D^* \ell \nu$. In this work we present a perturbative
matching calculation of the relevant current renormalizations 
to one-loop order. The numerical simulations for the above
physics analyses use MILC ensembles with improved glue and
$2+1$ Asqtad staggered sea quarks \cite{milc}. The light valence quarks are 
also generated from Asqtad staggered quarks and converted to
naive quarks. The heavy (charm and beauty) quarks are treated
with the Fermilab action. See Ref.~\cite{actions} for more
details on the actions and parameters used in the numerical 
simulations. 

\section{Definitions}

In this work we follow the analysis of Ref.~\cite{kron}, where the 
one-loop corrections to heavy-light and heavy-heavy current 
renormalizations were calculated for Fermilab heavy and Clover light 
quarks with Wilson glue. 

The heavy-light currents have the form
\begin{equation}
 J_{\mu}^{hl \, \rm lat} = \overline{\psi}_h \Gamma_{\mu} \psi_l \;,
\label{eq:curhl}
\end{equation}
where $\Gamma_{\mu} = \gamma_{\mu}$ or $\gamma_{\mu} \gamma_5$ and
$\psi_l$ denotes a naive Asqtad Dirac spinor. The Fermilab Dirac spinor,
$\psi_h$, is rotated by
\begin{equation}
 \psi_h = \psi \,[1 + a d_1 \mathbf{\gamma}\cdot \mathbf{D}]\;,
\end{equation} 
with the tree-level coefficient
\begin{equation}
 d_1 = \frac{1}{2 + m_0a} - \frac{1}{2(1 + m_0 a)}\;.
\end{equation} 
The heavy-heavy currents have the form
\begin{equation}
 J_{\mu}^{hh'\,\rm lat} = \overline{\psi}_h \Gamma_{\mu} \psi_{h'} \;,
\label{eq:curhh}
\end{equation}
where now both spinors are rotated Dirac spinors. 
Since the heavy quarks are rotated, the lattice currents of 
Eqns.~(\ref{eq:curhl}) and (\ref{eq:curhh}) include the leading order 
tree-level discretization corrections. 

The current renormalization is defined as
\begin{equation}
 Z^{hl}_{J_{\Gamma}} =  
 \frac{ (Z^{(1/2)}_{2h} \Lambda_{J_{\Gamma}} Z^{(1/2)}_{2l} )^{\rm cont}}
     { (Z^{(1/2)}_{2h} \Lambda_{J_{\Gamma}} Z^{(1/2)}_{2l} )^{\rm lat}} \;,
\label{eq:zhl}
\end{equation}
where $\Lambda_{J_{\Gamma}}$ are the vertex corrections and 
$Z_{2h}$ ($Z_{2l}$) are the heavy (light) quark wave function 
renormalizations. 

We factor out the dominant mass dependence due to the tree-level 
wave function renormalization of the
heavy Fermilab quark by defining the perturbative expansion as
\begin{equation}
 e^{-m_1^{[0]}a/2} Z^{hl}_{J_{\Gamma}} = 1 + g_0^{2}
                  Z^{hl\,[1]}_{J_{\Gamma}} + \ldots \;,
\end{equation}
where the heavy quark masses are defined as usual,
\begin{equation}
m_1^{[0]}a = \log (1+m_{0}a)\;,
\quad  m_{0}a = 1/(2\kappa_h) - 1/(2\kappa_{crit})\;. 
\end{equation}

Since $Z_{V_4}$ for degenerate masses is easy to calculate 
nonperturbatively, it is useful to define
\begin{equation}
 \rho^{hl}_{J_{\Gamma}} \equiv \frac{Z^{hl}_{J_{\Gamma}}}
               {\sqrt{Z^{hh}_{V_4} Z^{ll}_{V_4}}} 
= 1 + g_0^{2}
                  \rho^{[1]}_{J_{\Gamma}} + \ldots \;.
\label{eq:rho}
\end{equation}
In this case, the dominant mass dependence cancels by construction.

Analogously, for heavy-heavy currents we have:
\begin{equation}
 Z^{hh'}_{J_{\Gamma}} =  
 \frac{ (Z^{(1/2)}_{2h} \Lambda_{J_{\Gamma}} Z^{(1/2)}_{2h'} )^{\rm cont}}
     { (Z^{(1/2)}_{2h} \Lambda_{J_{\Gamma}} Z^{(1/2)}_{2h'} )^{\rm lat}} \;.
\label{eq:zhh}
\end{equation}
Taking the leading mass dependence out again, the perturbative
expansion is defined as 
\begin{equation}
 e^{-(m_{1h}^{[0]}+m_{1h'}^{[0]})a/2} Z^{hh'}_{J_{\Gamma}} = 1 + g_0^{2}
                  Z^{hh'\,[1]}_{J_{\Gamma}} + \ldots \;.
\end{equation}
Finally, the $\rho$ factors for heavy-heavy currents are defined as
\begin{equation}
 \rho^{hh'}_{J_{\Gamma}} \equiv \frac{Z^{hh'}_{J_{\Gamma}}}
               {\sqrt{Z^{hh}_{V_4} Z^{h'h'}_{V_4}}} 
     = 1 + g_0^{2} \rho^{hh'\,[1]}_{J_{\Gamma}} + \ldots \;.
\label{eq:rhohh}
\end{equation}

\section{Procedure}

In this work we use the automated perturbation theory techniques 
developed by L\"uscher and Weisz \cite{lw} to 
generate the Feynman rules for the lattice actions. We then
integrate the loop diagrams by ``brute-force'' using VEGAS 
\cite{vegas}.  The advantage of using automated perturbation theory
is that it is relatively easy to switch actions \cite{ht}. Indeed, we
have results for the current renormalizations for two gluon
actions, two light quark actions and the heavy quark action.

The one-loop diagrams for the vertex corrections (including
the rotations) are given in Ref.~\cite{kron}.
We have performed the following tests of our calculation:
\begin{itemize}

\item For the automated perturbation theory code, we have
compared our vertices and propagators against known results.

\item We have written two independent programs for calculating the
current renormalizations based on the automated perturbation theory
code.

\item We have a third independent calculation of the current 
renormalizations using traditional semi-analytic methods.

\item Our results for the heavy-heavy currents agree with those of 
Ref.~\cite{kron} when we switch from the improved gluon propagator to 
the Wilson gluon propagator. We also reproduce the results of 
Ref.~\cite{kron} for heavy-light currents with Clover light quarks
and Wilson glue.

\item Our result for the Asqtad naive wave function renormalization
agrees with Ref.~\cite{z2}, and our result for the naive-naive vertex
correction with Wilson glue agrees with Ref.~\cite{leesharpe}.

\end{itemize}

\vskip-0.2cm

\section{Results}

\begin{figure}[htb]
\centering
\includegraphics[clip=true,width=0.80\textwidth,angle=270]{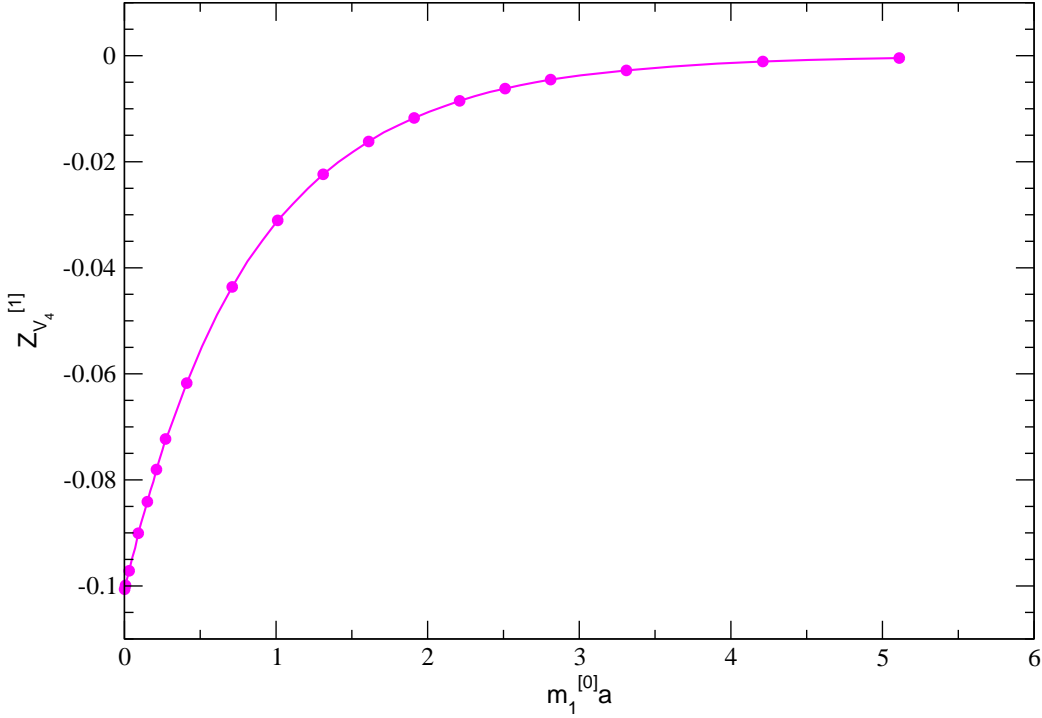}
\caption{$Z^{hh\,[1]}_{V_4}$ for equal masses as a function of $m^{[0]}_1$.}
 \end{figure}
\begin{figure}[htp]
\vskip-0.5cm
\centering
\includegraphics[clip=true,width=0.80\textwidth,angle=270]{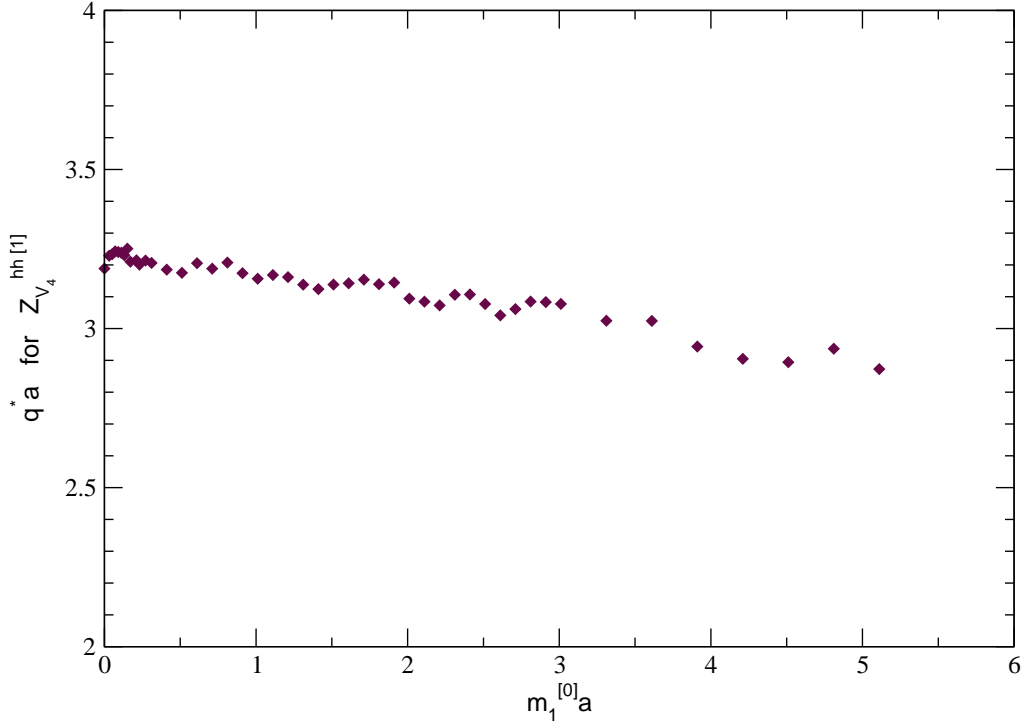}
\caption{$q^*a$ for $Z^{hh\,[1]}_{V_4}$ for equal masses as a function 
of $m^{[0]}_1$.}
 \end{figure}

Figures 1--4 illustrate our results for the $Z$'s, $\rho$'s, and
$q^*$'s as functions of the heavy-quark mass, $m_1^{[0]}a$. 
The $q^*$'s are calculated from the log moments using 
Eqn.~(13) of Ref.~\cite{hlm}.

Our results for the heavy-heavy currents are very similar to those
of Ref.~\cite{kron}, since they differ only in the gluon propagator. 
The main features of the mass dependence are the same. Figures~1--2
show results for the degenerate mass $V_4$ current. We also have results
for the other currents ($V_i$, $A_4$, $A_i$) as well as results for 
currents with unequal masses. In the massless limit we find
\begin{equation}
Z_{V_4}^{hh\,[1]} (m_1^{[0]} = 0) = -0.10056\,(3)\;,
\end{equation}
in good agreement with Ref.~\cite{ukawa}. This is another test of our
calculation.

Figure~3 shows a comparison of the current renormalization of the heavy-naive
$A_4$ current with the corresponding $\rho$ factor, and Figure~4 shows 
$\rho^{hl\, [1]}_{V_4}$ and $\rho^{hl\, [1]}_{V_i}$. First, the general
features of the heavy-quark mass dependence are similar to the 
results of Ref.~\cite{kron}. Second, $\rho^{hl\, [1]}_{A_4}$ is significantly
smaller than $Z^{hl\, [1]}_{A_4}$ over the relevant mass range. Hence, the 
cancellation between the numerator and denominator of Eq.~(\ref{eq:rho}) 
already observed in Ref.~\cite{kron} also takes place for heavy-naive
currents. 
\begin{figure}[htp]
\vskip-0.5cm
  \centering
 \includegraphics[clip=true,width=0.80\textwidth,angle=270]{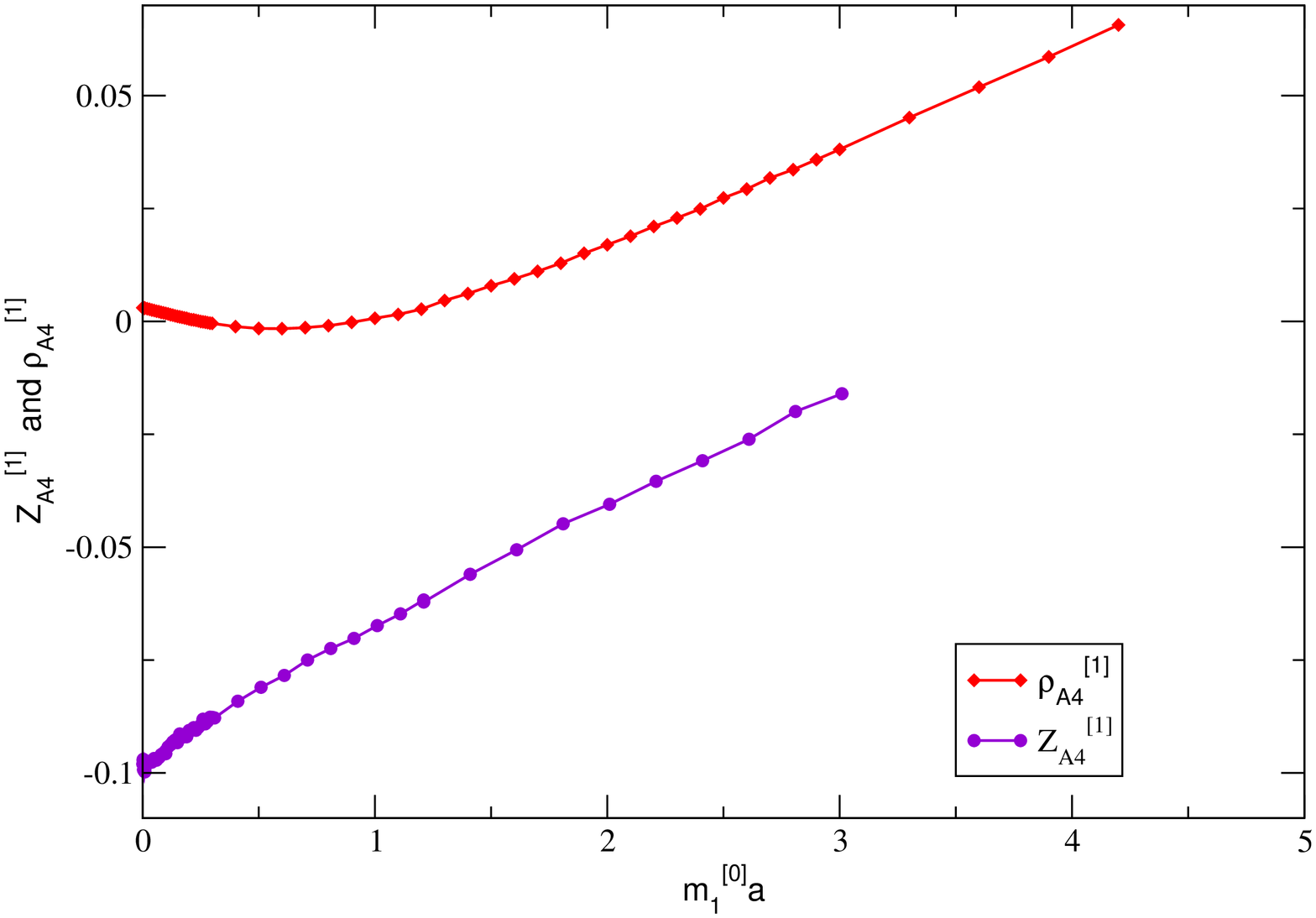}
  \caption{Comparison of $Z^{hl\, [1]}_{A_4}$ with $\rho^{hl}_{A_4}$ }
 \end{figure}
\begin{figure}[htb]
\vskip-0.5cm
  \centering
 \includegraphics[clip=true,width=0.80\textwidth,angle=270]{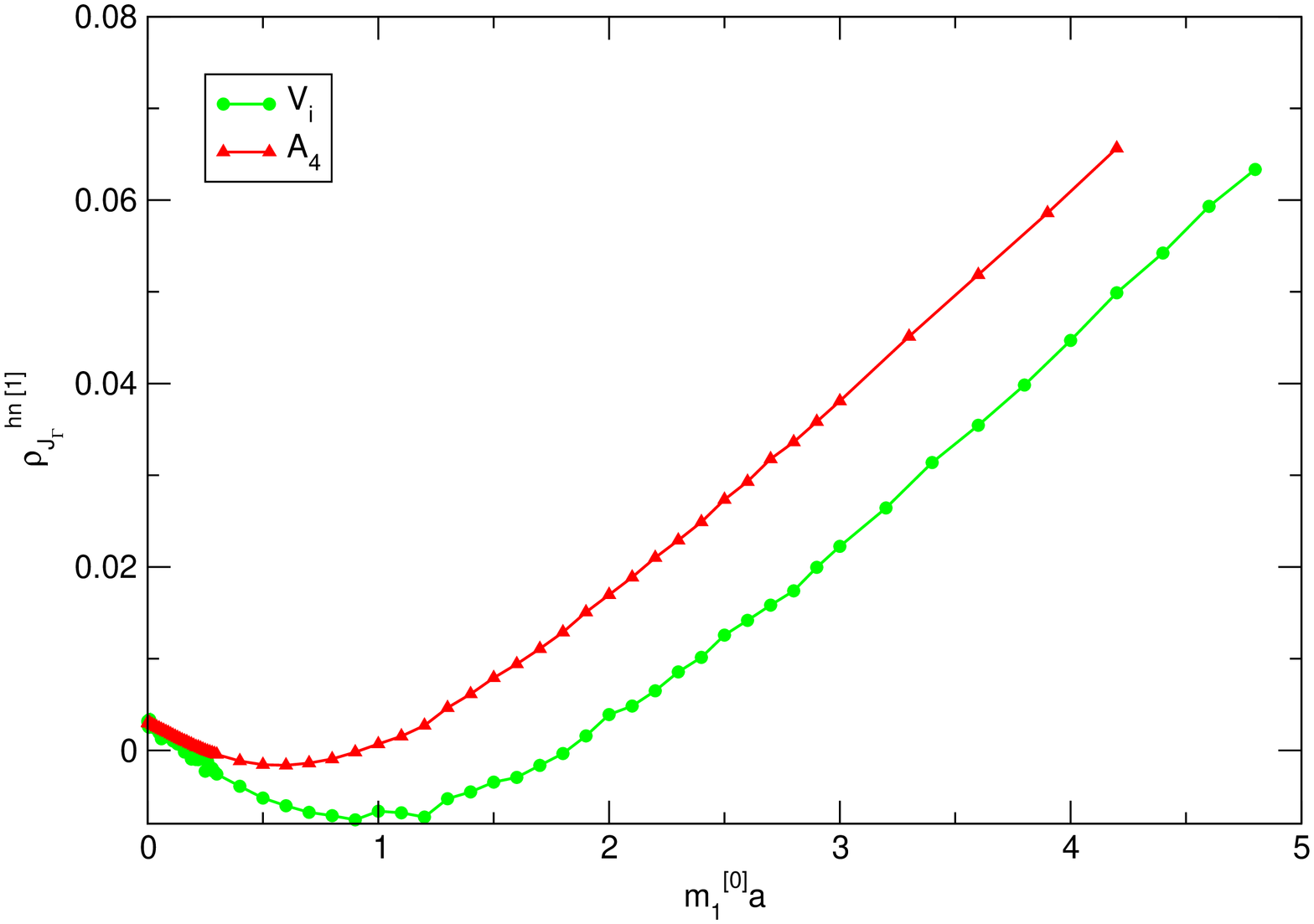}
  \caption{$\rho$ factors for heavy-naive currents}
 \end{figure}

In the massless limit we find:
\begin{eqnarray}
\rho_{V_4}^{hl\, [1]} (m_1^{[0]} = 0) & = & -3.038 \,(2)\cdot 10^{-3} 
\\ \nonumber
\rho_{V_i}^{hl\, [1]} (m_1^{[0]} = 0) & = & -3.05 \,(5) \cdot 10^{-3}
\end{eqnarray}

We also have results for the naive $V_4$ current renormalization. In the
massless limit we find
\begin{equation}
Z_{V_4}^{ll\, [1]} (m_0 = 0) = -0.10457\,(4).
\end{equation}
We have studied the mass dependence of $Z_{V_4}^{ll\,[1]}$ by varying $m_0$
between zero and the strange quark mass. We find that $Z_{V_4}^{ll\,[1]}$ 
is essentially independent of $m_0$.

In summary, we have calculated the current renormalizations relevant
for the numerical analyses of heavy-light decay constants and semileptonic
form factors performed by the Fermilab Lattice and MILC collaborations. 
We calculate the full mass dependence of the $Z$'s and $\rho$'s. The 
one-loop corrections to the $\rho$ factors are small. They vary 
roughly between $0.4$\% and 4\%, depending on lattice spacing. 

\section*{Acknowledgements}
We thank Howard Trottier for his help with testing M.N.'s
automated perturbation theory code, in particular by checking 
our Asqtad vertices and propagators.
This work was supported in part by the DOE 
under grant no. DE-FG02-91ER40677 and by the Junta de Andaluc\'{\i}a 
[P05-FQM-437 and P06-TIC-02302].
The numerical calculations for this work were carried out on the 
Fermilab lattice QCD clusters, which are a computing resource 
of the USQCD collaboration and are funded by the DOE. We are 
grateful to the Fermilab Computing Divisions for operating and 
maintaining the clusters. 
Fermilab is operated by Fermi Research Alliance, LLC, under Contract
No.~DE-AC02-07CH11359 with the United States Department of Energy.

\end{document}